%% file: template.tex
\documentclass[a4paper]{article}

\usepackage{INTERSPEECH2019}

\title{Neural Architecture Search For Keyword Spotting}
\name{Tong Mo$^{*1}$, Yakun Yu$^{*1}$, Mohammad Salameh$^{2}$, Di Niu$^{1}$, Shangling Jui$^{2}$}

\address{
  $^1$ University of Alberta, Edmonton, AB, Canada\\
  $^2$ Huawei Technologies}
\email{\{tmo1, yakun2, dniu\}@ualberta.ca \{mohammad.salameh, jui.shangling\}@huawei.com}

\begin{document}

\maketitle
\renewcommand{\thefootnote}{\fnsymbol{footnote}}
\setcounter{footnote}{-1}
\footnote{$^*$ Equal contributions, listed in alphabetical order.}
\begin{abstract}
   Deep neural networks have recently become a popular solution to keyword spotting systems, which enable the control of smart devices via voice. In this paper, we apply neural architecture search to search for convolutional neural network models that can help boost the performance of keyword spotting based on features extracted from acoustic signals while maintaining an acceptable memory footprint. Specifically, we use differentiable architecture search techniques to search for operators and their connections in a predefined cell search space. The found cells are then scaled up in both depth and width to achieve competitive performance. We evaluated the proposed method on Google's Speech Commands Dataset and achieved a state-of-the-art accuracy of over 97\% on the setting of 12-class utterance classification commonly reported in the literature. 
 
\end{abstract}

\section{Introduction}

Keyword spotting (KWS) aims to identify a set of keywords in utterances.
KWS was traditionally performed in the cloud based on audio recordings uploaded by users \cite{tang2018deep}.
Nowadays, on-device KWS applications are becoming increasingly popular,  e.g., Apple's ``Siri'', Microsoft's ``Cortana'' and Amazon's ``Alexa'', which help preserve user privacy and avoid data leakage during transmission. 
The deployment of KWS models on resource-constrained smart devices requires a small footprint while retaining accuracy. 
Thus, small-footprint KWS focuses on the recognition of simple commands, such as ``yes'', ``no'', ``on'' and ``off'', which are sufficient to support frequent user-device interactions.

In recent years, various convolutional neural networks (CNNs) have been applied to KWS and achieved remarkable results. Sainath et al. \cite{sainath2015convolutional} introduce CNNs into KWS and show that CNNs perform well when limiting the number of parameters. 
Tang et al. \cite{tang2018deep} use variants of the deep residual network (ResNet) to build a neural KWS model, and achieve an accuracy of 95.8\%  with 239K parameters on the Google Speech Commands Dataset (v1) \cite{warden2018speech} using Res15. 
Choi et al. \cite{choi2019temporal} combine temporal convolutions with ResNet to construct TC-ResNet models and improve the accuracy to 96.6\% with 305K parameters. 
Mittermaier et al. \cite{mittermaier2019small} use parameterized Sinc-convolutions from SincNet to classify keywords based on raw audio, and reduce the number of parameters to 122K while maintaining the accuracy of TC-ResNet.
Kao et al. \cite{kao2019sub} propose a sub-band CNN architecture to apply different convolutional kernels on each feature sub-band, and achieve an accuracy of around 90.0\% on the second version of Google Speech Commands Dataset while reducing the computation by 39.7\% compared to a full-band CNN model. Zeng et al. \cite{zeng2019effective} use DenseNet with BiLSTM and achieve an accuracy of 96.2\% following Google's setup \cite{sainath2015convolutional}.
Pons et al. \cite{pons2019randomly} propose a model that uses randomly weighted CNNs as feature extractors to conduct audio classification.
Chen et al. \cite{chen2019small} propose a compact and efficient convolutional network (CENet) for small-footprint KWS, and insert the graph convolutional network (GCN) for contextual feature augmentation to CENet as CENet-GCN, which can achieve an accuracy of 96.8\% with 72.3K parameters when only using Mel-frequency Cepstrum Coefficient (MFCC) features as the input.     
Majumdar et al. \cite{majumdar2020matchboxnet} propose MatchboxNet that contains residual blocks of 1D time-channel separable convolutions, batch-normalization (BN), ReLU, and dropout layers, achieving an accuracy of around 97.48\% with 93K parameters, though on a different setting of 30-class utterance classification with the help of data augmentation (while the majority of the literature evaluates a 12-class benchmark). 


In this paper, we propose to use Neural Architecture Search (NAS) to automate the neural network architecture design for KWS. NAS is widely used and evaluated for image classification and language modeling tasks.
Zoph et al. \cite{zoph2016neural} first use a reinforcement learning approach to train a neural network architecture with the maximum validation accuracy on CIFAR-10. However, this method is computationally expensive,  requiring hundreds of GPUs, and the model could not be transferred to a large dataset. 
The same authors then design a NASNet search space to search for the best convolutional layer (or ``cell") and stack copies of this cell to form a NASNet architecture \cite{zoph2018learning}. Though NASNet is trained faster and able to generalize to larger datasets, 
the whole search process still takes over four days with 500 GPUs. 
Other NAS methods, e.g., AmoebaNet \cite{real2019regularized}, Progressive NAS \cite{liu2018progressive}, have been proposed to further optimize the search process. However, all of them search over a discrete domain where more architecture evaluations are required. 
To make the search space continuous, Liu et al. \cite{liu2018darts} propose a differentiable architecture search (DARTS) and enable the efficient search of neural architectures through gradient descent.

To date, there have been some efforts on NAS for KWS, although not achieving state-of-the-art results. Veniat et al. \cite{veniat2019stochastic} propose a stochastic adaptive neural architecture search approach for KWS that automatically adapts the architecture by a recurrent neural network (RNN) according to the difficulty of the prediction problem at each time step, and achieve an 86.5\% prediction accuracy on the Google Speech Commands Dataset \cite{warden2018speech}. Anderson et al. \cite{anderson2020performance} propose a performance-oriented neural architecture search approach based on information about the hardware and achieve a 95.11\% prediction accuracy on the same dataset.

In this paper, we leverage DARTS \cite{liu2018darts}, a gradient-based differentiable NAS technique to search for the best convolutional network architecture for KWS. 
A typical NAS process involves searching for the best architecture for a given task, followed by training the found best architecture from scratch. The search process involves three dimensions \cite{elsken2018neural}. 
The \textit{search space} defines which architectures are considered and the operations that compose them.
\textit{Search strategies} define the strategy used to explore the search space, e.g., reinforcement learning (RL) \cite{zoph2016neural,zoph2018learning, tan2018mnasnet,pham2018efficient}, evolutionary algorithm \cite{elsken2018efficient, real2017large,real2019regularized} and gradient-based approaches \cite{dong2019searching,liu2018darts, chen2019progressive, xie2018SNAS}. 
It is computationally intensive to evaluate the proposed architecture by the search strategy from scratch. \textit{Performance estimation} estimates the performance of an architecture without the need to train it fully.
Research in NAS aims to improve in these dimensions in order to discover highly performing architectures while minimizing the search cost (in terms of GPU days). 

We choose a differentiable NAS approach, DARTS, because it has remarkable efficiency, as compared to earlier NAS techniques operated in a discrete domain based on RNN controllers \cite{zoph2016neural,zoph2018learning}. DARTS can finish searching in a single GPU day.  
Besides, DARTS does not rely on performance predictors \cite{liu2018progressive} and can find architectures with complex structures in a rich search space.

We evaluate the proposed NAS method on the public Google Speech Commands Dataset \cite{warden2018speech}. 
Our experimental results have shown that the proposed method can find architectures that achieve a state-of-the-art accuracy of over 97\% on the common benchmark setting of 12-class utterance classification, which is the same evaluation setting adopted by most KWS literature \cite{tang2017honk, tang2018deep, choi2019temporal,chen2019small,mittermaier2019small,veniat2019stochastic}.



\begin{figure}[t]
    \centering
    \includegraphics[width = \columnwidth]{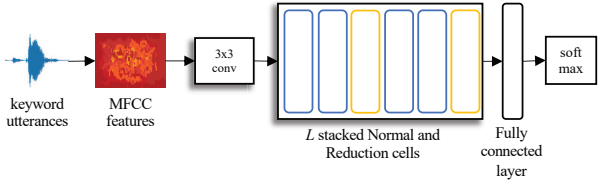}
    \vspace{-6.5mm}
    \caption{The composition of the convolutional neural network to be searched for. A stack of six cells is used during search, where each blue rectangle represents a normal cell, while each yellow one represents a reduction cell. Once the best cells are found, the network can be scaled up in both depth and width.}
    \label{fig:network}
\end{figure}

\section{Method}
We search for a convolutional neural network (CNN) to optimize the classification performance based on a matrix of MFCC features extracted from each audio sample. 
 As is shown in Figure~\ref{fig:network}, the CNN we will search for is composed of a head layer that performs a preliminary $3\times 3$ convolution, followed by a sequence of $L$ stacked layers, each called a \textit{cell}, and finally, a stem that performs the classification.  The preprocessing procedure to process audio in MFCC features will be described in Section~\ref{sec:experiment}. 
 To reduce the complexity of the search, we search for cell architectures rather than searching for the entire network architecture.
 As is illustrated in Figure~\ref{fig:network}, two types of cells are searched for: \textit{normal cells} and \textit{reduction cells}. A normal cell ensures that the size of its output is the same as that of its input by using a stride of one. A reduction cell, on the other hand, doubles the number of channels and divides the height and width of its input by one half. All the normal cells share the same neural architecture. So do all the reduction cells. Once the best normal cell and reduction cell architectures are found from the search phase, we will scale up the depth and width of the network by stacking the found cells and tuning the number of channels at the initial layer. When stacking cells sequentially to form a deeper architecture, the same stacking rule applies--every two normal cells are followed by a reduction cell.
 
\begin{figure}[t]
    \centering
    \includegraphics[width = \columnwidth]{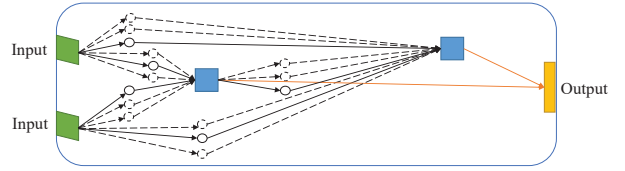}
    \vspace{-6.5mm}
    \caption{An illustration of the inner search space of a cell. Each circle is an operation in $\mathcal{O}$, while the solid ones are those finally selected by the search algorithm.}
    \label{fig:cell}
\end{figure}

We leverage a cost-efficient differentiable architecture search algorithm, DARTS \cite{liu2018darts}, to find the best normal and reduction cell architectures for KWS.
Specifically, a cell can be represented by a directed acyclic graph (DAG) consisting of ordered nodes and directed edges, as is shown in Figure~\ref{fig:cell}. There are two inputs to a cell (green), which correspond to the outputs of the previous two cells, while the output of the cell (yellow) is a concatenation of all intermediate nodes. 

Each node is a \textit{latent representation}, while each edge comprises mixed operations from a predefined operation set $\mathcal{O}$, e.g., $3\times3$ convolution, $5\times5$ convolution, max-pooling, etc. 
A directed edge connecting node $i$ and node $j$ represents the direction of information flow and performs a weighted sum $f_{i,j}(x_i)$ of all possible operations $o(.)\in \mathcal{O}$ applied onto the latent representation $x_i$ of node $i$,  i.e., 
\[
f_{i,j}(x_i) = \sum_{o \in \mathcal{O}}^{} \alpha_{(i,j), o} o(x_i).
\]
Let $\alpha^{(i,j)}$ denote the vector of $\alpha_{(i,j), o}$'s for all $o(.)\in \mathcal{O}$ on edge $(i,j)$.
The weights $\alpha^{(i,j)}$'s are learnable parameters, which encode the cell structure.
The latent representation $x_j$ for each intermediate node $j$ is then computed as the sum of outputs from all its preceding nodes, i.e., 
\[
x_j = \sum_{i < j}^{} f_{i, j}(x_i).
\]

For simplicity, let $\alpha$ denote the \textit{architecture weights}, i.e., the vector concatenating all $\alpha^{(i,j)}$'s on all the edges, and let $w$ denote the model weights of the corresponding architecture. 
Denote the training loss by $\mathcal{L}_{train}$ and the validation loss by $ \mathcal{L}_{val}$.
The DARTS algorithm searches for the best architecture (encoded by $\alpha$) by solving a bi-level optimization problem: 
\begin{equation}
\label{eqn:alphaop}
\begin{aligned}
\min_{\alpha}&\quad \mathcal{L}_{val}(w^*(\alpha), \alpha) \\
\mbox{s.t.} &\quad w^*(\alpha) = \arg\min_w \mathcal{L}_{train}(w,\alpha),
\end{aligned}
\end{equation}
where $\alpha$ and $w$ are the upper level and lower level parameters, respectively.
The goal is to find $\alpha^*$ that minimizes the validation loss $\mathcal{L}_{val}(w^*(\alpha), \alpha)$ such that $w^*$ under the given architecture weights $\alpha$ is obtained by minimizing the training loss $\mathcal{L}_{train}(w,\alpha)$. Architecture weights $\alpha$ and model weights $w$ are learned jointly using gradient descent \cite{liu2018darts} until convergence: 1) updating the architecture weights $\alpha$ by descending based on $\nabla_\alpha \mathcal{L}_{val}(w , \alpha)$; 2) updating the neural network weights $w$ by descending  based on $\nabla_w\mathcal{L}_{train}(w,\alpha)$.
      
At the end of the search, the operation $o$ with the highest weight $\alpha_{(i,j), o}$ on edge $(i,j)$ will be finally selected, as illustrated in Figure~\ref{fig:cell} by the solid circles. Only the selected operations and the edges connected to them are kept to produce the resulting cell architecture. 



%

\section{Performance Evaluation}
\label{sec:experiment}

We evaluate the proposed method for keyword spotting on Google Speech Commands Dataset (v1) \cite{warden2018speech}. This dataset contains 65,000 one-second-long audio utterances pertaining to 30 words. There are approximately 2,200 samples for each word. Following the same setting as \cite{tang2017honk,tan2018mnasnet,choi2019temporal,chen2019small,mittermaier2019small}, we cast the problem as a classification task that distinguishes among 12 classes, i.e., ``yes'', ``no'', ``up'', ``down'', ``left'', ``right'', ``on'', ``off'', ``go'', ``stop'', an unknown class, and a silence class. The unknown class contains utterances sampled from the remaining 20 words other than the above ten words, while the silence class has utterances with only background noise. We split the entire dataset into 40\% training, 40\% validation, and 20\% testing sets. The training set and validation set are used during architecture search, and are further combined to form a new training set for evaluating the best architecture on the test set.

\input{Table3}

\subsection{Experimental setup}


We follow the preprocessing procedure of Honk \cite{tang2017honk} to process the acoustic signals, which are adopted by multiple small-footprint KWS studies \cite{tang2017honk,tang2018deep,sainath2015convolutional,choi2019temporal,chen2019small}. To generate training data, we first add background noise to each sample with 80\% probability at each epoch, followed by a random $t$-second time shift where $t$ is sampled from a UNIFORM$[-100, 100]$ distribution on each sample to enhance robustness. Then, we apply a 20Hz/4kHz filter. 
Finally, each raw audio file is split into 101 frames using a window size of 30 milliseconds and a frameshift of 10 milliseconds. We extract 40 Mel-Frequency cepstral coefficients (MFCC) features for each frame and stack them across the time axis.

During neural architecture search, we set the number of cells to 6 and train the network for 50 epochs.
The batch size and the initial number of channels are both set to 16 to ensure that the network fits into one GPU. 
We use stochastic gradient descent (SGD) to update the weights $\omega$ with a momentum of 0.9 and a weight decay of $3 \times 10^{-4}$. The learning rate for $\omega$ is set to 0.025, following a cosine annealing scheduler.
We optimize the architecture parameters $\alpha$ with Adam ($\beta_1=0.5$, $\beta_2=0.999$), and set weight decay and the initial learning rate to  $10^{-3}$ and $3 \times 10^{-4}$, respectively.

During the evaluation, we instantiate the network to be tested based on the best cell architecture with the highest validation score found by the search phase, and experiment with a depth of 6 and 12. A network of depth 6 is illustrated in Fig.~\ref{fig:network}. A network of depth 12 is obtained by stacking the 6-cell network twice. 
We randomly re-initialize the weights in the network and re-train it from scratch for 200 epochs to report the evaluation results.

In our cell searches, each normal/reduction cell consists of 7 nodes.
Table~\ref{tab:opSets} summarizes the candidate operations used.
In total, 7 candidate operations have been considered : skip connection (or identity), \textit{zero}, average pooling, max pooling, dilated convolution,  separable convolution (depthwise separable convolution), and regular convolution. \textit{Zero} means no connection between two nodes, identity represents identity mapping. 
The dilated convolution introduces a dilation rate (set to two in our experiments) to the regular convolution.
Each convolution operation follows the sequence of execution: ReLU, Convolution, Batch Normalization (BN). Each separable convolution executes two ReLU-Conv-BN sequences.

As shown in Table~\ref{tab:opSets}, we conduct searches based on two sets of operators. \textit{NAS1} uses separable convolutions, dilated convolutions and pooling, while \textit{NAS2} uses regular convolutions instead of separable convolutions. 
The separable convolution consists of a depth-wise convolution conducted independently over each channel of the input, followed by a point-wise convolution, i.e., a $1 \times 1$ convolution, to combine information across channels \cite{chollet2017xception,chen2018encoder}. 
Dilated convolutions are known to be able to expand the receptive field exponentially without loss of coverage \cite{yu2015multi}, while separable convolutions can reduce the number of parameters and computational cost \cite{kaiser2017depthwise}.  
Separable convolutions are also frequently used in neural KWS literature \cite{mittermaier2019small,majumdar2020matchboxnet} to improve performance and reduce model size.

On the other hand, NAS2 uses the regular convolution, which is the convolutional operation traditionally used in ResNet and has been applied to KWS by \cite{tang2018deep}.
NAS2 considers the same operations used in \cite{tang2018deep} to test whether our search strategy is effective at producing architectures that can beat traditional ResNet models \cite{tang2018deep} when using similar operations. 



\begin{figure}[!t]
    \centering
    \includegraphics[width = \columnwidth]{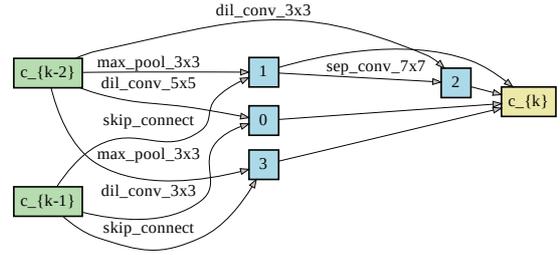}
    \vspace{-6.5mm}
    \caption{The normal cell found on the NAS1 search space.}
    \label{fig:normal_plan3}
\end{figure}

\begin{figure}[!t]
    \centering
    \includegraphics[width = \columnwidth]{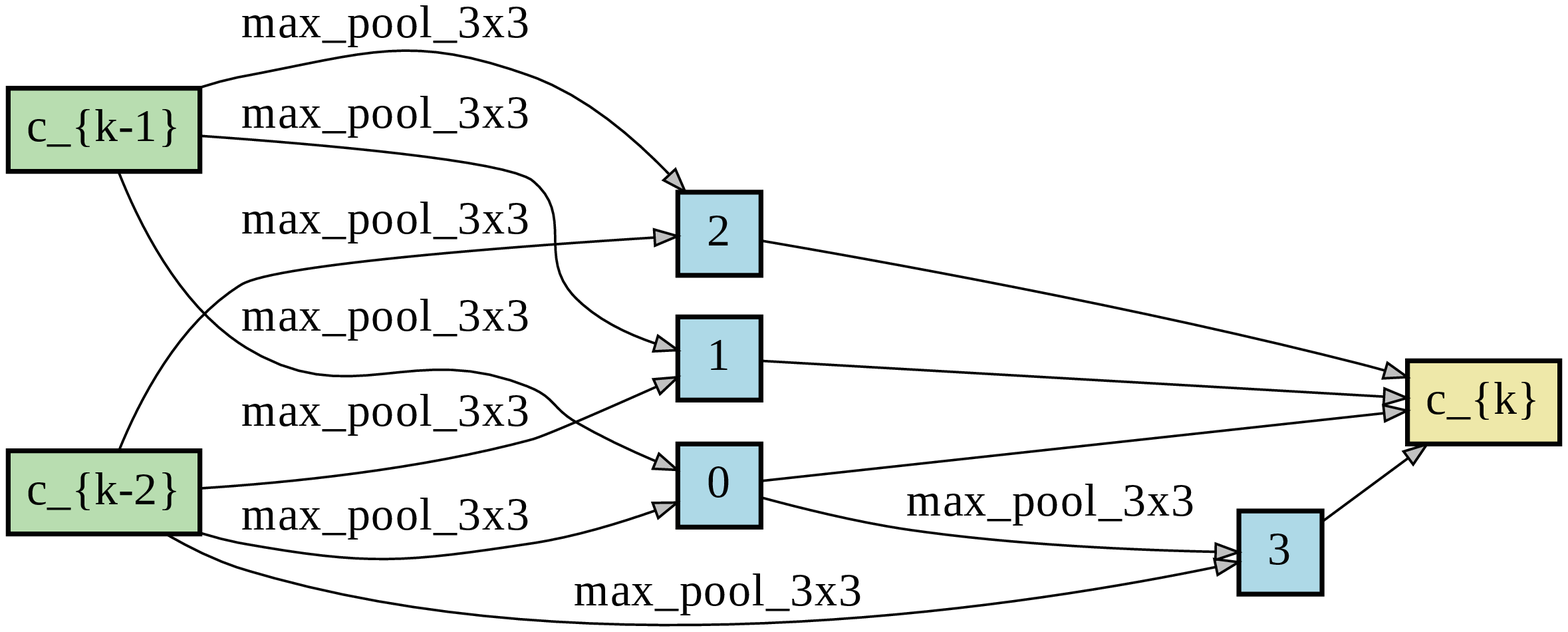}
    \vspace{-6.5mm}
    \caption{The reduction cell found on the NAS1 search space.}
    \label{fig:reduction_plan3}
\end{figure}

\begin{figure}[!t]
    \centering
    \includegraphics[width = \columnwidth]{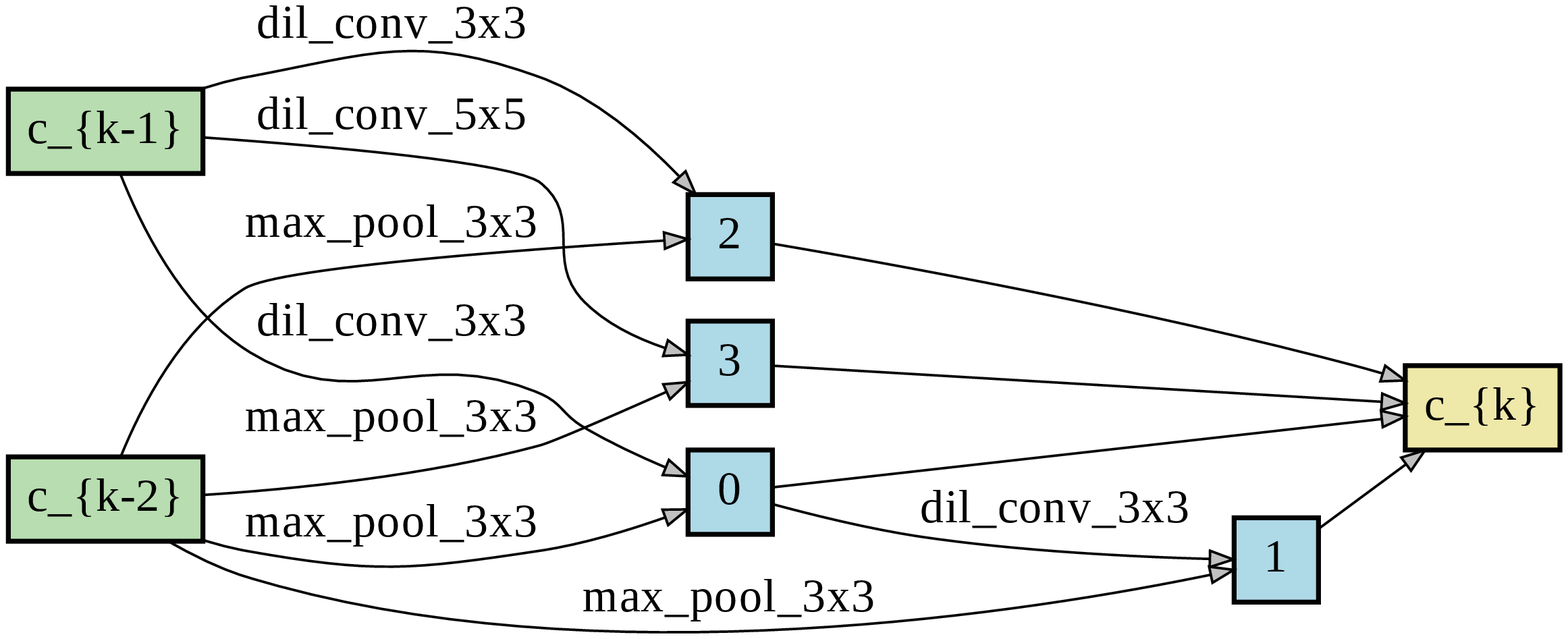}
    \vspace{-6.5mm}
    \caption{The normal cell found on the NAS2 search space.}
    \label{fig:normal_honk}
\end{figure}

\begin{figure}[!t]
    \centering
    \includegraphics[width = \columnwidth]{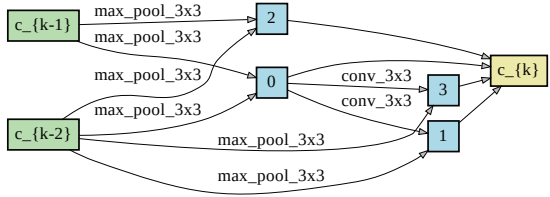}
    \vspace{-10.5mm}
    \caption{The reduction cell found on the NAS2 search space.}
    \label{fig:reduction_honk}
\end{figure}

We evaluate the models discovered under both NAS1 and NAS2 search spaces in terms of the accuracy and model size, under different scaling-up settings, by varying the depth (number of cells) and the initial number of channels.  
We compare to the following baseline models that 
 utilize CNN blocks and 
are evaluated on the same dataset and 12 classes as our method \footnote{MatchboxNet-$3\times2\times64$ \cite{majumdar2020matchboxnet} proposes a deep residual network and achieves state-of-the-art results on Google Speech Dataset v1 on 30 keyword classes. Thus, their setup is not comparable to ours or the listed baselines. It also uses data augmentation, e.g., time shift perturbations and SpecAugment to boost the performance, which is not used in our method or the listed baselines. 
Similarly, we do not compare to DenseNet-BiLSTM \cite{zeng2019effective} which relies on attention BiLSTM.
}:
\begin{itemize}
    \item Res15: a ResNet variant based on regular convolutions achieving the highest accuracy in \cite{tang2018deep}. It consists of 6 residual blocks and 45 feature maps. 
    \item TC-ResNet14-1.5: a ResNet variant achieving the highest accuracy in \cite{choi2019temporal}, which uses a $3 \times 1$ \textit{temporal convolution} instead of regular convolutions to reduce the footprint. 6 residual blocks are used. A width multiplier of 1.5 is applied to expand the number of channels at each layer.
    \item SincConv+DSConv: the best model reported in \cite{mittermaier2019small}, which first uses the Sinc-convolution to extract features from raw audio and then applies separable convolutions with a kernel length of 9 to reduce the model size.
    \item CENet-GCN-40: the best model in \cite{chen2019small}, which mainly consists of bottleneck blocks and a GCN module. Each bottleneck block is a stack of $1\times1$, $3\times3$ and $1\times1$ convolutions to reduce model complexity. The GCN module is introduced to learn non-local relations of convolutional features. 
\end{itemize}

\subsection{Results}

Figures~\ref{fig:normal_plan3}-\ref{fig:reduction_honk} illustrate the cells found on each search space. The search costs for NAS1 and NAS2 remain at a low level of 0.58 GPU day and 0.29 GPU day, respectively. Table 2 shows a performance comparison between our models and baseline models. 
From this table, we can observe that the model found by NAS1 with 6 cells and 16 initial channels outperform Res15, TC-ResNet and SincConv in terms of both accuracy and the number of parameters, while the rest of the NAS1 models can achieve an accuracy higher than CENet-GCN-40. Notably, NAS1 with 12 cells and 16 channels could achieve a new state-of-the-art accuracy of 97.06\% with an acceptable model size of 281K parameters, under the setting of 12 classes on the same dataset.

For NAS2 models, comparing with Res15 \cite{tang2018deep}, which uses a similar operation space. It is worth noting that NAS2 only uses the operations that appear in Res15, and does not use any temporal convolutions or separable convolutions, thus could lead to a fairly large model size.
However, they all achieve a better accuracy of over 96.7\%, outperforming Res15,
while the accuracy of NAS2 with 6 cells and 16 initial channels is 0.94 percentage points higher than Res15 with a model size 24\% smaller than that of Res15. The results of this set of experiments demonstrate the benefits and necessity of architecture search even under the same operation space. Although using the same set of operations, architectures with better performance can be found with NAS.

Moreover, we investigate the impact of the depth by changing the number of cells, and the impact of the width by changing the number of initial channels. From NAS1 and NAS2, we observe that the model performance can be improved by increasing the depth or width, although at a cost of an increased model size. 
In addition, NAS1 models tend to have fewer parameters than NAS2 models due to the use of separable convolutions.

\input{Table2}
\vspace{-5.5pt}

\section{Conclusion}

Existing methods for neural keyword spotting rely on manually designed convolutional neural networks and other neural networks. In this paper, we perform differentiable neural architecture search to search for CNN architectures that can lead to a high accuracy and a relatively small footprint. Our approach is robust and finds architectures with accuracy over 96\% under different sets of operations. 
The best models found by neural architecture search achieves a state-of-the-art accuracy of over 97\% accuracy on the Google Speech Commands Dataset, outperforming a range of existing baseline models under the same experimental setup, while maintaining competitive footprints. These observations demonstrate the enormous potential of conducting neural architecture search for keyword spotting, especially toward other types of neural networks and the adoption of KWS-friendly operations, which open up avenues for future investigation.



\bibliographystyle{IEEEtran}

\bibliography{template}

\end{document}

%% file: Table3.tex

\begin{table}[t]
\centering
\caption{The candidate operations used.}
\label{tab:opSets}
\begin{tabular}{lccc}
\hline
\textbf{Model}  & \textbf{Search space} \\ \hline
\begin{tabular}[c]{@{}c@{}c@{}c@{}} NAS1 \end{tabular}   & 
\begin{tabular}[l]{@{}l@{}l@{}l@{}} \textit{zero}, $3\times3$ max\_pool, $3\times3$  avg\_pool, identity,\\ $3\times3$ and  $5\times5$ dil\_conv, $5\times5$, $7\times7$,  and \\$9\times9$  sep\_conv \end{tabular}    \\ \hline

\begin{tabular}[c]{@{}c@{}c@{}c@{}} NAS2 \end{tabular}   & 
\begin{tabular}[l]{@{}l@{}l@{}l@{}} \textit{zero}, $3\times3$ max\_pool, $3\times3$ avg\_pool, identity,\\ $3\times3$ and  $5\times5$ dil\_conv, $3\times3$ regular\_conv \end{tabular}   \\ \hline

\end{tabular}

\end{table}

%% file: Table2.tex
\begin{table}[!t]
\centering
\caption{Performance of the models found by the proposed method and baseline models. The numbers marked with $\dagger$ are taken from the corresponding papers. '-' means not available. The best results among different methods are marked in bold.}

\begin{tabular}{lcccc}

\hline

\textbf{Model}    
&
 \textbf{\begin{tabular}[c]{@{}c@{}} Cell \\(\#)\end{tabular}}
  &
 \textbf{\begin{tabular}[c]{@{}c@{}} Channels \\(\#)\end{tabular}}
 &
 \textbf{\begin{tabular}[c]{@{}c@{}} Acc.\\ (\%)\end{tabular}}    & \textbf{\begin{tabular}[c]{@{}c@{}} Par.\\(K)\end{tabular}}
\\\hline
Res15 \cite{tang2018deep}  & - & -         & 95.8$\dagger$                                                & 239     \\ 
TC-ResNet14-1.5 \cite{choi2019temporal} & - & -   & 96.6$\dagger$                                                & 305       \\
SincConv+DSConv \cite{mittermaier2019small}                 & - & -         & 96.6$\dagger$                      & 122     \\ 

CENet-GCN-40 \cite{chen2019small}   & - & -         & 96.8$\dagger$                      & 72.3     \\ 
\hline

NAS1   & 
\begin{tabular}[c]{@{}c@{}c@{}c@{}} 6 \\ 6 \\ 6 \\ 12 \end{tabular} & 
\begin{tabular}[c]{@{}c@{}c@{}c@{}} 16 \\ 24 \\ 36 \\ 16 \end{tabular}     & 
\begin{tabular}[c]{@{}c@{}c@{}c@{}} 96.74 \\ 96.90 \\ 96.96 \\ \textbf{97.06} \end{tabular}     & 
\begin{tabular}[c]{@{}c@{}c@{}c@{}} \textbf{107} \\ 223 \\ 474 \\ 281 \end{tabular}  \\   
\hline

 NAS2   & 
\begin{tabular}[c]{@{}c@{}c@{}c@{}} 6 \\ 6 \\ 6 \\ 12 \end{tabular} & 
\begin{tabular}[c]{@{}c@{}c@{}c@{}} 16 \\ 24 \\ 36 \\ 16 \end{tabular}     & 
\begin{tabular}[c]{@{}c@{}c@{}c@{}} 96.74 \\ 96.86 \\ \textbf{97.22} \\ 96.81 \end{tabular}     & 
\begin{tabular}[c]{@{}c@{}c@{}c@{}} \textbf{182} \\ 400 \\ 886 \\ 281 \end{tabular} \\
\hline
\end{tabular}

\end{table}

%% file: template.bbl
\begin{thebibliography}{10}
\providecommand{\url}[1]{#1}
\csname url@samestyle\endcsname
\providecommand{\newblock}{\relax}
\providecommand{\bibinfo}[2]{#2}
\providecommand{\BIBentrySTDinterwordspacing}{\spaceskip=0pt\relax}
\providecommand{\BIBentryALTinterwordstretchfactor}{4}
\providecommand{\BIBentryALTinterwordspacing}{\spaceskip=\fontdimen2\font plus
\BIBentryALTinterwordstretchfactor\fontdimen3\font minus
  \fontdimen4\font\relax}
\providecommand{\BIBforeignlanguage}[2]{{%
\expandafter\ifx\csname l@#1\endcsname\relax
\typeout{** WARNING: IEEEtran.bst: No hyphenation pattern has been}%
\typeout{** loaded for the language `#1'. Using the pattern for}%
\typeout{** the default language instead.}%
\else
\language=\csname l@#1\endcsname
\fi
#2}}
\providecommand{\BIBdecl}{\relax}
\BIBdecl

\bibitem{tang2018deep}
R.~Tang and J.~Lin, ``Deep residual learning for small-footprint keyword
  spotting,'' in \emph{2018 IEEE International Conference on Acoustics, Speech
  and Signal Processing (ICASSP)}.\hskip 1em plus 0.5em minus 0.4em\relax IEEE,
  2018, pp. 5484--5488.

\bibitem{sainath2015convolutional}
T.~N. Sainath and C.~Parada, ``Convolutional neural networks for
  small-footprint keyword spotting,'' in \emph{Sixteenth Annual Conference of
  the International Speech Communication Association}, 2015.

\bibitem{warden2018speech}
P.~Warden, ``Speech commands: A dataset for limited-vocabulary speech
  recognition,'' \emph{arXiv preprint arXiv:1804.03209}, 2018.

\bibitem{choi2019temporal}
S.~Choi, S.~Seo, B.~Shin, H.~Byun, M.~Kersner, B.~Kim, D.~Kim, and S.~Ha,
  ``Temporal convolution for real-time keyword spotting on mobile devices,''
  \emph{arXiv preprint arXiv:1904.03814}, 2019.

\bibitem{mittermaier2019small}
S.~Mittermaier, L.~K{\"u}rzinger, B.~Waschneck, and G.~Rigoll,
  ``Small-footprint keyword spotting on raw audio data with
  sinc-convolutions,'' \emph{arXiv preprint arXiv:1911.02086}, 2019.

\bibitem{kao2019sub}
C.-C. Kao, M.~Sun, Y.~Gao, S.~Vitaladevuni, and C.~Wang, ``Sub-band
  convolutional neural networks for small-footprint spoken term
  classification,'' \emph{arXiv preprint arXiv:1907.01448}, 2019.

\bibitem{zeng2019effective}
M.~Zeng and N.~Xiao, ``Effective combination of densenet and bilstm for keyword
  spotting,'' \emph{IEEE Access}, vol.~7, pp. 10\,767--10\,775, 2019.

\bibitem{pons2019randomly}
J.~Pons and X.~Serra, ``Randomly weighted cnns for (music) audio
  classification,'' in \emph{ICASSP 2019-2019 IEEE International Conference on
  Acoustics, Speech and Signal Processing (ICASSP)}.\hskip 1em plus 0.5em minus
  0.4em\relax IEEE, 2019, pp. 336--340.

\bibitem{chen2019small}
X.~Chen, S.~Yin, D.~Song, P.~Ouyang, L.~Liu, and S.~Wei, ``Small-footprint
  keyword spotting with graph convolutional network,'' \emph{arXiv preprint
  arXiv:1912.05124}, 2019.

\bibitem{majumdar2020matchboxnet}
S.~Majumdar and B.~Ginsburg, ``Matchboxnet--1d time-channel separable
  convolutional neural network architecture for speech commands recognition,''
  \emph{arXiv preprint arXiv:2004.08531}, 2020.

\bibitem{zoph2016neural}
B.~Zoph and Q.~V. Le, ``Neural architecture search with reinforcement
  learning,'' \emph{arXiv preprint arXiv:1611.01578}, 2016.

\bibitem{zoph2018learning}
B.~Zoph, V.~Vasudevan, J.~Shlens, and Q.~V. Le, ``Learning transferable
  architectures for scalable image recognition,'' in \emph{Proceedings of the
  IEEE conference on computer vision and pattern recognition}, 2018, pp.
  8697--8710.

\bibitem{real2019regularized}
E.~Real, A.~Aggarwal, Y.~Huang, and Q.~V. Le, ``Regularized evolution for image
  classifier architecture search,'' in \emph{Proceedings of the aaai conference
  on artificial intelligence}, vol.~33, 2019, pp. 4780--4789.

\bibitem{liu2018progressive}
C.~Liu, B.~Zoph, M.~Neumann, J.~Shlens, W.~Hua, L.-J. Li, L.~Fei-Fei,
  A.~Yuille, J.~Huang, and K.~Murphy, ``Progressive neural architecture
  search,'' in \emph{Proceedings of the European Conference on Computer Vision
  (ECCV)}, 2018, pp. 19--34.

\bibitem{liu2018darts}
H.~Liu, K.~Simonyan, and Y.~Yang, ``Darts: Differentiable architecture
  search,'' \emph{arXiv preprint arXiv:1806.09055}, 2018.

\bibitem{veniat2019stochastic}
T.~V{\'e}niat, O.~Schwander, and L.~Denoyer, ``Stochastic adaptive neural
  architecture search for keyword spotting,'' in \emph{ICASSP 2019-2019 IEEE
  International Conference on Acoustics, Speech and Signal Processing
  (ICASSP)}.\hskip 1em plus 0.5em minus 0.4em\relax IEEE, 2019, pp. 2842--2846.

\bibitem{anderson2020performance}
A.~Anderson, J.~Su, R.~Dahyot, and D.~Gregg, ``Performance-oriented neural
  architecture search,'' \emph{arXiv preprint arXiv:2001.02976}, 2020.

\bibitem{elsken2018neural}
T.~Elsken, J.~H. Metzen, and F.~Hutter, ``Neural architecture search: A
  survey,'' \emph{arXiv preprint arXiv:1808.05377}, 2018.

\bibitem{tan2018mnasnet}
M.~Tan, B.~Chen, R.~Pang, V.~Vasudevan, M.~Sandler, A.~Howard, and Q.~V. Le,
  ``Mnasnet: Platform-aware neural architecture search for mobile,'' 2018.

\bibitem{pham2018efficient}
H.~Pham, M.~Y. Guan, B.~Zoph, Q.~V. Le, and J.~Dean, ``Efficient neural
  architecture search via parameter sharing,'' \emph{arXiv preprint
  arXiv:1802.03268}, 2018.

\bibitem{elsken2018efficient}
T.~Elsken, J.~H. Metzen, and F.~Hutter, ``Efficient multi-objective neural
  architecture search via lamarckian evolution,'' \emph{arXiv preprint
  arXiv:1804.09081}, 2018.

\bibitem{real2017large}
E.~Real, S.~Moore, A.~Selle, S.~Saxena, Y.~L. Suematsu, J.~Tan, Q.~V. Le, and
  A.~Kurakin, ``Large-scale evolution of image classifiers,'' in
  \emph{Proceedings of the 34th International Conference on Machine
  Learning-Volume 70}.\hskip 1em plus 0.5em minus 0.4em\relax JMLR. org, 2017,
  pp. 2902--2911.

\bibitem{dong2019searching}
X.~Dong and Y.~Yang, ``Searching for a robust neural architecture in four gpu
  hours,'' in \emph{Proceedings of the IEEE Conference on Computer Vision and
  Pattern Recognition}, 2019, pp. 1761--1770.

\bibitem{chen2019progressive}
X.~Chen, L.~Xie, J.~Wu, and Q.~Tian, ``Progressive differentiable architecture
  search: Bridging the depth gap between search and evaluation,'' 2019.

\bibitem{xie2018SNAS}
\BIBentryALTinterwordspacing
S.~Xie, H.~Zheng, C.~Liu, and L.~Lin, ``{SNAS:} stochastic neural architecture
  search,'' \emph{CoRR}, vol. abs/1812.09926, 2018. [Online]. Available:
  \url{http://arxiv.org/abs/1812.09926}
\BIBentrySTDinterwordspacing

\bibitem{tang2017honk}
R.~Tang and J.~Lin, ``Honk: A pytorch reimplementation of convolutional neural
  networks for keyword spotting,'' \emph{arXiv preprint arXiv:1710.06554},
  2017.

\bibitem{chollet2017xception}
F.~Chollet, ``Xception: Deep learning with depthwise separable convolutions,''
  in \emph{Proceedings of the IEEE conference on computer vision and pattern
  recognition}, 2017, pp. 1251--1258.

\bibitem{chen2018encoder}
L.-C. Chen, Y.~Zhu, G.~Papandreou, F.~Schroff, and H.~Adam, ``Encoder-decoder
  with atrous separable convolution for semantic image segmentation,'' in
  \emph{Proceedings of the European conference on computer vision (ECCV)},
  2018, pp. 801--818.

\bibitem{yu2015multi}
F.~Yu and V.~Koltun, ``Multi-scale context aggregation by dilated
  convolutions,'' \emph{arXiv preprint arXiv:1511.07122}, 2015.

\bibitem{kaiser2017depthwise}
L.~Kaiser, A.~N. Gomez, and F.~Chollet, ``Depthwise separable convolutions for
  neural machine translation,'' \emph{arXiv preprint arXiv:1706.03059}, 2017.

\end{thebibliography}
